\documentclass[12pt,preprint]{aastex}      

\usepackage[latin1]{inputenc}
\usepackage{lscape}

\begin{document}

\title{A Magnetized Jet from a Massive Protostar}

\author{Carlos~Carrasco-Gonz\'alez\altaffilmark{1,2}, Luis~F.~Rodr\'{\i}guez\altaffilmark{2}, Guillem~Anglada\altaffilmark{1}, Josep Mart\'{i}\altaffilmark{3}, Jos\'e M. Torrelles\altaffilmark{4}, Mayra Osorio\altaffilmark{1}}

\altaffiltext{1}{Instituto Astrof\'{\i}sica Andaluc\'{\i}a, CSIC, Camino Bajo de Hu\'etor 50, E-18008 Granada, Spain; $^2$Centro de Radioastronom\'{\i}a y Astrof\'{\i}sica UNAM, Apartado Postal 3-72 (Xangari), 58089 Morelia, Michoac\'an, Mexico; $^3$Departamento de F\'{\i}sica, EPSJ, Universidad de Ja\'en, Campus Las Lagunillas s/n, Edif. A3, 23071 Ja\'en, Spain; $^4$Instituto de Ciencias del Espacio (CSIC/IEEC)-UB, Universitat de Barcelona, Mart\'{\i} i Franqu\`{e}s 1, E-08028 Barcelona, Spain}

\textbf{Synchrotron emission is commonly found in relativistic jets from active galactic nuclei (AGNs) and microquasars, but so far its presence in jets from young stellar objects (YSOs) has not been proved. Here, we present evidence of polarized synchrotron emission arising from the jet of a YSO. The apparent magnetic field, with strength of $\sim$0.2 milligauss, is parallel to the jet axis, and the polarization degree increases towards the jet edges, as expected for a confining helical magnetic field configuration. These characteristics are similar to those found in AGN jets, hinting at a common origin of all astrophysical jets.}

\vspace{1cm}

 Supersonic jets are observed to emerge from a wide variety of astrophysical systems, from young brown dwarfs to active galatic nuclei (AGNs). Despite their different physical scales (from hundreds to billions of astronomical units), they have strong morphological similarites (\emph{1}), and a common feature of these systems is the presence of a gaseous disk around the central object. However, it is yet unclear whether there is a universal mechanism that can explain the origin of all these jets.

It is believed that the formation of stars takes place with the assistance of an accretion disk that transports gas and dust from the envelope of the system to the protostar and a jet that removes angular momentum and magnetic flux, allowing the accretion to proceed (\emph{2}). Theoretical models for the formation of jets from young stellar objects (YSOs) assume that the initial acceleration takes place relatively close to the protostar as gas from the accretion disk is lifted and centrifugally accelerated along magnetic field lines. This mechanism was first suggested as the origin of jets from accretion disks around black holes (\emph{3,4}) and was soon proposed as the mechanism for protostellar jets (\emph{5,6}). Present observational techniques cannot resolve and directly probe into the innermost regions where the jet is launched, but theoretical models for protostellar jets are able to explain some bulk observational properties, such as the correlation between accretion and ejection rates (\emph{7}). The collimation of the outflow into a jet takes place at larger distances from the star and is also believed from theoretical arguments to take place magnetically. The magnetic field lines anchored in the rotating accretion disk are proposed to wrap in a helical configuration that is needed to provide the magnetic pressure to collimate the jet via ``hoop" stresses (\emph{8,9}). Observations aimed at measuring the magnetic field in protostellar jets are very scarce and imply complex modeling and/or a large number of assumptions (\emph{10,11}).

 A very powerful method to simultaneously obtain the structure and strength of the magnetic field in jets is through observations of the synchrotron emission. When electrons move at relativistic velocities in the presence of a magnetic field, they emit linearly polarized synchrotron radiation. The spectrum of the synchrotron emission in the centimeter wavelength range is characterized by a negative spectral index (i.e., the flux density decreases with frequency), and it is related to the strength of the magnetic field. Additionally, the direction of the polarization is related to the direction of the magnetic field. This method has been the most successful in the study of the magnetic field in jets from AGNs (\emph{12}) and microquasars (\emph{13}), where the ejected material moves at relativistic velocities. In contrast, protostellar jets, where the bulk of the material moves at much smaller velocities (200 to 1000 km~s$^{-1}$) (\emph{14,15}), apparently would lack relativistic electrons that can provide information on the magnetic field. Then, the emission is usually dominated by a different mechanism (free-free emission originated from the thermal motions of the electrons, characterized by a positive spectral index and no polarization). Intringuingly, in a few protostellar jets (associated with both low- and high-mass YSOs) radio emission with negative spectral indices, which could correspond to nonthermal synchrotron emission from relativistic electrons, has been reported [Serpens (\emph{16}), HH~80-81 (\emph{17}), Cepheus A (\emph{18}), W3(H$_2$O) (\emph{19}), L778 VLA 5 (\emph{20}), IRAS~16547$-$4247 (\emph{21})] (\emph{22}). If relativistic electrons were present, the acceleration of particles would be most likely produced where the fast thermal jet impacts on the surrounding medium and a strong shock wave is formed. Thus, particles might be accelerated up to relativistic energies by the Fermi mechanism (\emph{23}). In order to confirm the presence of synchrotron radiation in these protostellar jets, a clear detection of linearly polarized radio emission is required. Because polarized emission is only a fraction of the continuum emission and protostellar jets are usually very weak, highly sensitive observations are required.
 
 We obtained highly sensitive (total observation time $\sim$12 hours) radio continuum images of the HH~80-81 jet using the Very Large Array (VLA) in its C configuration (\emph{24}). This highly collimated radio jet is driven by the protostellar source IRAS 18162$-$2048, whose total luminosity is $\sim$17,000~$L_\odot$ (where $L_\odot$ is the luminosity of the Sun) [if this luminosity originated in a single object, it would correspond to a zero-age main-sequence star of $\sim$10 $M_\odot$ (where $M_\odot$ is the mass of the Sun), according to the stellar evolutionary models of (\emph{25})]. The radio jet consists of a chain of radio sources aligned in a southwest-northeast direction. To the southwest, it terminates in the optical and radio Herbig-Haro objects HH~80 and HH~81, whereas to the northeast it terminates in the radio Herbig-Haro object HH~80~N (\emph{17}) (Fig. 1). With a total extension of 5.3 pc [for an assumed  distance of 1.7~kpc (\emph{26})], this is one of the largest and most collimated protostellar jets known so far. Previous radio observations showed that the spectrum of the emission from the central source is characterized by a positive spectral index, suggesting that it is dominated by thermal free-free emission (\emph{17}). In contrast, the negative spectral indices of the emission from HH~80-81, HH~80~N, as well as from some of the knots in the jet lobes, suggest that an additional nonthermal component could be present in these sources (\emph{17}).

 Our observations (Figure 1) show that the emission of the knots located $\sim$0.5~pc from the driving source is linearly polarized, indicating the presence of nonthermal synchrotron emission in this jet (\emph{24}) and implying the presence of relativistic electrons and a magnetic field. The observed polarization vectors are perpendicular to the direction of the jet, with a degree of polarization of the order of 10 to 30\%. The direction of the apparent magnetic field (the component in the plane of the sky averaged over the line of sight) is taken to be perpendicular to the polarization vectors [this is correct for a nonrelativistically moving source, such as the HH~80-81 jet; for relativistic jets, such as AGN jets, additional assumptions on the velocity field are required (\emph{27})]. Then, the apparent magnetic field appears very well aligned with the direction of the HH~80-81 radio jet (Fig. 1C).
 
 By using the equations 7.14 and 7.15 of (\emph{28)} we obtained the minimum total energy, $E=c_{13}(1+k)^{4/7}\phi^{3/7}R^{9/7}L_{\rm R}^{4/7}$, and the equipartition magnetic field, $B=[4.5c_{12}(1+k)/\phi]^{2/7}R^{-6/7}L_{\rm R}^{2/7}$. In these equations, $\phi$ is the filling factor of the emitting region, $R$ is the radius of the source, $L_{\rm R}$ is the integrated radio luminosity, $k$ is the ratio between the heavy particle energy and the electron energy, and $c_{12}$ and $c_{13}$ are functions of the spectral index and of the minimum and maximum frequencies considered in the integration of the spectrum, which are given in appendix 2 of (\emph{28}). To estimate the radio luminosity, we integrated the radio spectrum in the range between 20 and 2 cm by using the flux density measurements of (\emph{17}) (table S1). We used a filling factor of $\phi$=0.5 and $k$=40 [this value of $k$ is appropriate for acceleration in a nonrelativistic shock (\emph{29})]. With these parameters, we obtained typical values for the knots of the radio jets of $B\simeq 0.2$~mG and $E\simeq2\times10^{43}$~erg.

 The direction of the apparent magnetic field obtained from the synchrotron emission is strongly concentrated in a direction very close ($\sim4^\circ$) to the jet axis, whereas the direction of the magnetic field inferred from polarization of the dust near the protostar (\emph{30}) is offset by $\sim20^\circ$ and shows more scatter (Figure 2); however, part of this scatter could be attributed to turbulent and thermal motions. It has been estimated that the magnetic field in the region $\sim$0.1~pc around the protostar IRAS~18162$-$2048 has a value of 0.2~mG (\emph{30}). If we assume that this field is anchored to the dust envelope and behaves like a dipole, it should drop with distance cubed. However, even at $\sim$0.5~pc from the protostar, the strength of the magnetic field in the jet remains comparable to that observed in the core by (\emph{30}), when it should have decreased by more than a factor of 100 if it was merely reflecting the field in the ambient cloud. These results suggest that, whereas dust polarization traces the magnetic field associated with the ambient material close to the protostar, the synchrotron emission traces the magnetic field intrinsically associated with the jet.

 The polarization properties and the magnetic field configuration in the HH~80-81 jet are very similar to those observed in AGN jets. In AGN jets, the polarization is always observed either perpendicular or parallel to the axis of the jet (\emph{27}). When jets are transversally resolved, the degree of linear polarization reaches a minimum towards the jet axis and increases towards the jet edges [e.g., Figure 4a in \emph{(31)}]. Figure 1B shows that the linear polarization in HH~80-81 is perpendicular to the jet axis, and Figure 3 shows that the degree of linear polarization increases as a function of the distance from the jet axis. In AGN jets, this configuration of the polarization has been interpreted as indicative of a large-scale helical magnetic field (\emph{27,31}).

The relevance of magnetic fields in protostellar jets has been anticipated from theoretical models of star formation [see (\emph{4,32,33}) and references therein]. However, most of the attention has been paid to the role of the magnetic field in the launching region, and its relevance in the collimation and propagation at larger scales has been scarcely studied, even theoretically (\emph{34}). Possible signatures of magnetic fields on the large scales of the YSO outflows have been searched with multidimensional numerical MHD simulations of overdense, radiative cooling jets [see (\emph{35}) and references therein]. These studies find that the effects of magnetic fields on the outflow are dependent on both the field geometry and the strength, which so far have been poorly determined from observations [e.g., (\emph{35})]. Our results for the HH 80-81 jet provide a measure of the strength and structure of the magnetic field in a YSO jet at large distances ($\sim$0.5 pc) from the driving source. Jets from YSOs produce thermal line emission from shocks that in turn provides  direct, independent information on their physical conditions (density, temperature and velocity) (\emph{36}) not available for jets from AGNs and microquasars. Our results in the HH 80-81 jet suggest that mapping of linear polarization in a set of protostellar jets (associated with both low- and high-mass YSOs) in combination with detailed theoretical modeling may lead to a deeper understanding of the overall jet phenomenon and of the relevance of magnetic fields in star formation. Although these radio jets are usually weak, these studies will be feasible with the next generation of ultrasensitive radio interferometers. Recent molecular maser observations are starting to provide information on the strength and direction of the magnetic field in the disks around massive protostars (\emph{37}); observations of the dust polarization provide information on the strength and direction of the magnetic field in the envelope feeding the disk in the protostellar system (\emph{30,38}). Thus, we may be on the brink of describing the magnetic field characteristics in the complete envelope-disk-outflow system with a combination of observational techniques.

 Furthermore, recent theoretical works [e.g., (\emph{39})] suggest that protostellar jets could be a source of gamma rays. These models postulate, as a working hypothesis, the presence of relativistic electrons in such jets. Our detection of synchrotron emission in the HH~80-81 jet demonstrates the presence of relativistic electrons in the jets from massive protostars, providing an observational ground to the theoretical conjecture and making these objects a potential target for future high-energy studies.

\vspace{1cm}

\begin{center}
\textbf{REFERENCES AND NOTES}
\end{center}

\begin{enumerate}

\item{} De Young, D.S., Astrophysical jets. \emph{Science.} \textbf{252}, 389-396 (1991)

\item{} This scenario is well stablished in the case of low-mass stars, while for high-mass stars the evidence of disk/jet systems is more scarce [e.g., (\emph{40})].

\item{} Blandford R.D., Payne, D.G., Hydromagnetic flows from accretion discs and the production of radio jets. \emph{Mon. Not. R. Astron. Soc.} \textbf{199}, 883-903 (1982).

\item{} McKinney, J.~C., Blandford, R.~D., Stability of relativistic jets from rotating, accreting black holes via fully three-dimensional magnetohydrodynamic simulations. {\em Mon. Not. R. Astron. Soc.} {\bf 394}, L126-L130 (2009).

\item{} Pudritz, R.E., Norman, C.A., Centrifugally driven winds from contracting molecular disks. \emph{Astrophys. J.} \textbf{274}, 677-697 (1983) 

\item{} Shu, F. et al., Magnetocentrifugally driven flows from young stars and disks. 1: A generalized model. \emph{Astrophys. J.} \textbf{429}, 781-796 (1994)

\item{} Ferreira, J., Magnetically-driven jets from Keplerian accretion discs. {\em Astron. Astrophys.} {\bf 319}, 340-359 (1997).

\item{} Livio, M., The formation of astrophysical jets, in {\em Accretion Phenomena and Related Outflows, IAU Colloq.~163.} ASP Conference Series, Vol. 121 (eds D.T. Wickramasinghe, D.T., Ferrario, L. \& Bicknell, G.V.) 845-866 (Astronomical Society of the Pacific, San Francisco, 1997).

\item{} Pudritz, R.E., Ouyed, R., Fendt, Ch., Brandenburg, A., Disk winds, jets and outflows: theoretical and computational foundations, in {\em Protostars \& Planets V} (eds Reipurth, B., Jewitt, D., Keil, K.) 277-294 (University of Arizona Press, Tucson, 2007).

\item{} Chrysostomou, A., Lucas, P.~W., Hough, J.~H., Circular polarimetry reveals helical magnetic fields in the young stellar object HH135-136.\ {\em Nature} {\bf 450}, 71-73 (2007).

\item{} Morse, J.A., Heathcote, S., Cecil, G., Hartigan, P., Raymond, J.C. The bow shock and Mach disk of HH 111V.\ {\em Astrophys. J.} {\bf 410}, 764-776 (1993).

\item{} Bridle, A.H., Perley, R.A., Extragalactic radio jets. \emph{Annu. Rev. Astron. Astrophys.} \textbf{22}, 319-358 (1984)

\item{} Mirabel, I.F., Rodr\'{i}guez, L.F., Sources of Relativistic Jets in the Galaxy. \emph{Annu. Rev. Astron. Astrophys.} \textbf{37}, 409-443 (1999)

\item{} Mart\'{i}, J., Rodr\'{i}guez, L.F., Reipurth, B., Large proper motions and ejection of new condensations in the HH 80-81 thermal radio jet. \emph{Astrophys. J.} \textbf{449}, 184 (1995)

\item{} Pech, G. et al., Confirmation of a recent bipolar ejection in the very young hierarchical multiple system IRAS 16293$-$2422. \emph{Astrophys. J.} \textbf{712}, 1403-1409 (2010)

\item{} Rodr\'{i}guez, L.F. et al., Large proper motions in the remarkable triple radio source in Serpens. \emph{Astrophys. J.} \textbf{346}, L85-L88 (1989)

\item{} Mart\'{i}, J., Rodr\'{i}guez, L.F., Reipurth, B., HH 80-81: A highly collimated Herbig-Haro complex powered by a massive young star. \emph{Astrophys. J.} \textbf{416}, 208 (1993)

\item{} Garay, G. et al., The nature of the radio sources within the Cepheus A star-forming region. \emph{Astrophys. J.} \textbf{459}, 193 (1996)

\item{} Wilner, D.J., Reid, M.J., Menten, K.M., The synchrotron jet from the H$_2$O maser source in W3(OH). \emph{Astrophys. J.} \textbf{513}, 775-779 (1999)

\item{} Girart, J.M, Curiel, S., Rodr\'{i}guez, L.F., Cant\'o, J. Radio Continuum Observations towards Optical and Molecular Outflows. \emph{Rev. Mex. Astron. Astrof.} \textbf{38}, 169-186 (2002)

\item{} Rodr\'{i}guez, L.F., Garay, G., Brooks, K., Mardones, D., High angular resolution observations of the collimated jet source associated with a massive protostar in IRAS 16547$-$4247. \emph{Astrophys. J.} \textbf{626}, 953-958 (2005)

\item{} Circularly polarized emission has been reported in the ejecta of the young star T Tauri South (\emph{41}), but in this case the proposed mechanism is gyrosynchrotron, that has different polarization characteristics than synchrotron and is expected to be significant only close to the stellar surfaces.

\item{} Bell, A.R., The acceleration of cosmic rays in shock fronts. I.  \emph{Mon. Not. R. Astron. Soc.} \textbf{182}, 147-156 (1978)

\item{} See supporting material on Science Online.

\item{} Schaller, G., Schaerer, D., Meynet, G., Maeder, A. New grids of stellar models from 0.8 to 120 $M_\odot$ at Z = 0.020 and Z = 0.001, {\it Astron. Astrophys. Suppl. Ser.}, {\bf 96,} 269.

\item{} Rodr\'{i}guez, L.F., Moran, J.M., Ho, P.T.P., Gottlieb, E.W., Radio observations of water vapor, hydroxyl, silicon monoxide, ammonia, carbon monoxide, and compact H II regions in the vicinities of suspected Herbig-Haro objects. \emph{Astrophys. J.} \textbf{235}, 845-865 (1980)

\item{} Lyutikov, M., Pariev, V.I., Gabuzda, D.C., Polarization and structure of relativistic parsec-scale AGN jets. \emph{Mon. Not. R. Astron. Soc.} \textbf{360}, 869-891 (2005)

\item{} Pacholczyk, A.G., Radio Astrophysics (San Francisco: Freeman, 1970)

\item{} Beck, R., Krause, M., Revised equipartition and minimum energy formula for magnetic field strength estimates from radio synchrotron observations. \emph{Astron. Nachr.} \textbf{326}, 414-427 (2005)

\item{} Curran, R. L., Chrysostomou, A., Magnetic fields in massive star-forming regions. \emph{Mon. Not. R. Astron. Soc.} \textbf{382}, 699-716 (2007)

\item{} G\'omez, J.L. et al., Faraday rotation and polarization gradients in the jet of 3C~120: interaction with the external medium and a helical magnetic field?. \emph{Astrophys. J.} \textbf{681}, L69-L72 (2008)

\item{} Cabrit, S., The accretion-ejection connexion in T Tauri stars: jet models vs. observations, in \emph{Star-Disk Interaction in Young Stars.} Proceedings IAU Symp. 243, (eds Bouvier, J. \& Appenzeller, I.) 203-214 (Cambridge: Cambridge Univ. Press, 2007).

\item{} Lizano, S., Shu, F.~H., Galli, D., Glassgold, A., Magnetized disks around young stars, in {\em Magnetic Fields in the Universe II: From Laboratory and Stars to the Primordial Universe.} Rev. Mex. Astron. Astrof. (Ser. Conf.), Vol. 36 (eds Esquivel, A., Franco, J., Garc\'{\i}a-Segura, G., de Gouveia Dal Pino, E.M., Lazarian, A., Lizano, S. \& Raga, A.) 149-154 (Instituto de Astronom\'{\i}a, UNAM, Mexico City, 2009).

\item{} Ray, T.~P., Getting to grips with the unknown: how important are magnetic fields in outflows from young stars?, in {\em Magnetic Fields in the Universe II: From Laboratory and Stars to the Primordial Universe.} Rev. Mex. Astron. Astrof. (Ser. Conf.), Vol. 36 (eds Esquivel, A., Franco, J., Garc\'{\i}a-Segura, G., de Gouveia Dal Pino, E.M., Lazarian, A., Lizano, S. \& Raga, A.) 179-185 (Instituto de Astronom\'{\i}a, UNAM, Mexico City, 2009).

\item{} de Gouveia dal Pino, E.M., The role of magnetic fields on astrophysical jets, in \emph{Magnetic Fields in the Universe: From Laboratory and Stars to Primordial Structures.} AIP Conference Proceedings, Vol. 784 (eds de Gouveia dal Pino, E.M., Lugones, G. \& Lazarian, A.) 183-194 (American Institute of Physics, Melville, New York, 2005).

\item{} Reipurth, B., Bally, J., Herbig-Haro Flows: Probes of early stellar evolution.\ {\em Ann. Rev. Astron.  Astrophys.} {\bf 39}, 403-455 (2001).

\item{} Vlemmings, W.H.T., Surcis, G., Torstensson, K.J.E., van Langevelde, H.J., Magnetic field regulated infall on the disc around the massive protostar Cepheus A HW2. \emph{Mon. Not. R. Astron. Soc.} \textbf{404}, 134 (2010).

\item{} Girart, J.M., Rao, R., Marrone, D.P., Magnetic fields in the formation of sun-like stars. \emph{Science} \textbf{313}, 812 (2006)

\item{} Bosch-Ramon, V., Romero, G.E., Araudo, A.T., Paredes, J.M., Massive protostars as gamma-ray sources. \emph{Astron. Astrophys.} \textbf{511}, 8 (2010)

\item{} Patel et al., A disk of dust and molecular gas around a high-mass protostar. {\it Nature.} {\bf 437,} 109-111 (2005).

\item{} Ray et al., Large-scale magnetic fields in the outflow from the young stellar object T Tauri S. {\it Nature.} {\bf 385,} 415-417 (1997).

\item{} NRAO is a facility of the National Science Foundation, operated under cooperative agreement by Associated Universities, Inc. We acknowledge I. Agudo, A. Alberdi and J.L. G\'omez for useful comments. G.A., C.C.-G., J.M., M.O., and J.M.T. acknowledge support from MICINN (Spain) AYA2008-06189-C03 and AYA2007-68034-C03-02 grants (co-funded with FEDER funds), and from Junta de Andaluc\'{i}a (Spain). L.F.R. acknowledges the support of DGAPA, UNAM, and of CONACyT, M\'exico.

\end{enumerate}

\newpage












\begin{figure}  
\begin{center}

\includegraphics[height=0.74\textheight]{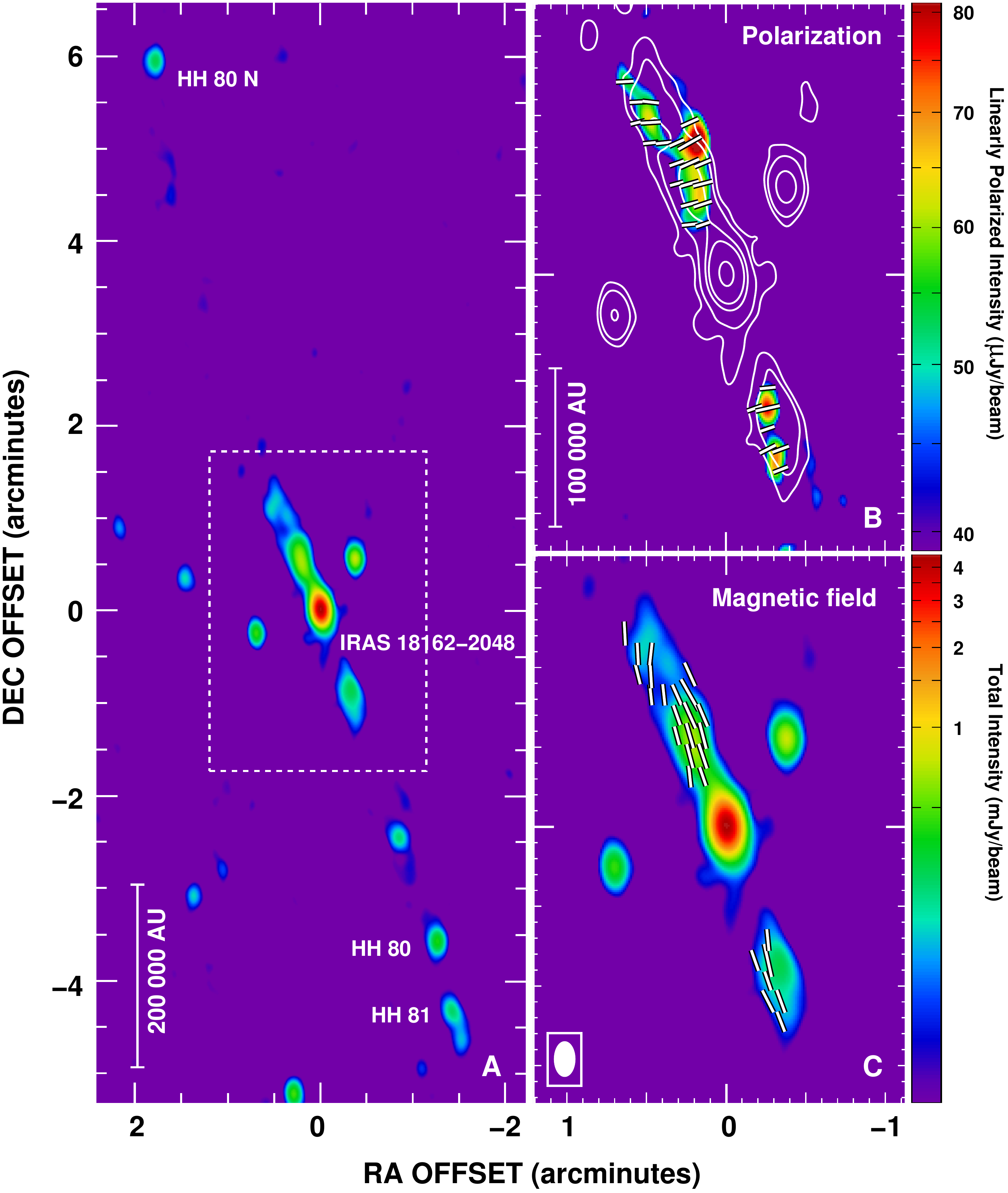}  

\caption{\footnotesize{Images of the HH~80-81 jet region at 6~cm wavelength. \textbf{(A)} Image of the total continuum intensity, showing the whole extension of the HH~80-81 jet. The brightest radio knot (labeled as IRAS~18162$-$2048) is associated with the central protostar. The color scale [shown in panel (C)] ranges from 0.039 to 4.5 mJy~beam$^{-1}$. The rectangle marks the central region of the jet, which is shown in (B) and (C). \textbf{(B)} Linearly polarized continuum intensity image (color scale). The color scale ranges from 39 to 85 $\mu$Jy~beam$^{-1}$. Polarization direction is shown as white bars. The total continuum intensity is also shown (contours). Contour levels are 40, 100, 400, 850, and 3300 $\mu$Jy~beam$^{-1}$. \textbf{(C)} The apparent magnetic field direction (white bars) is superposed with the total continuum intensity image (color scale). The images were made by using a tapering of 20 k$\lambda$ in order to emphasize extended emission. The root mean square of the noise is 0.013 mJy~beam$^{-1}$, and the synthesized beam is $13"$ by $8"$ with a position angle of 2$^\circ$ [shown in the bottom left corner of panel (C)]. The (0,0) offset position corresponds to the phase center of the observations, at right ascension (RA, J2000) = $18^h \, 19^m \, 12.102^s$ and declination (DE, J2000) = $-20^\circ \, 47' \, 30.61"$.}} 

\label{Fig1}  
\end{center} 
\end{figure}

\begin{landscape}

\begin{figure}  
\begin{center}

\includegraphics[height=0.7\textwidth]{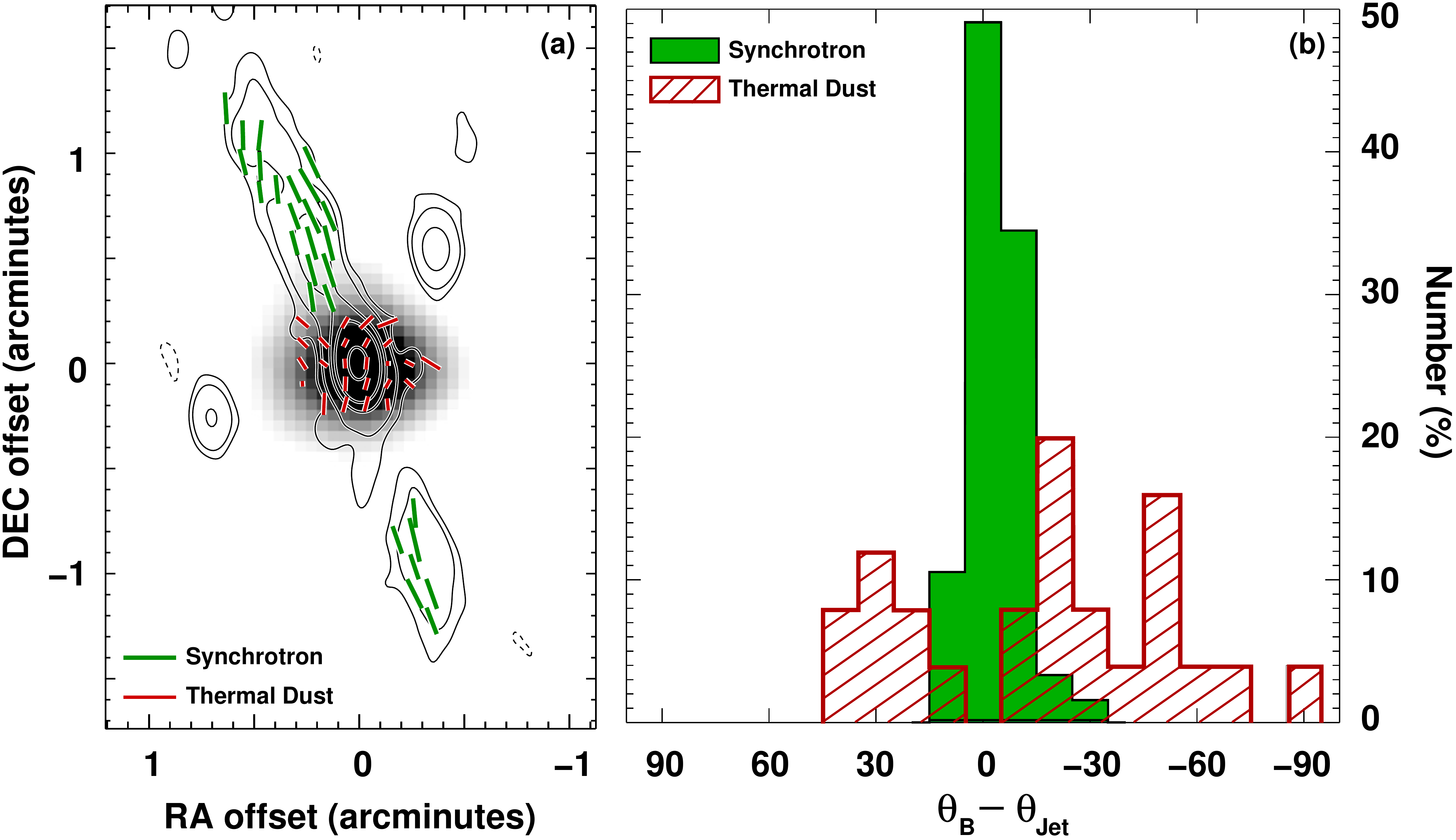}  

\caption{\footnotesize{\textbf{(A)} Image of the continuum emission of the HH~80-81 jet at 6~cm (contours) superposed with the dust continuum emission at 850~$\mu$m [gray scale (\emph{30})]. Green bars mark the direction of the apparent magnetic field inferred from the polarized synchrotron emission detected in our VLA observations. Red bars mark the direction of the magnetic field inferred from the dust polarization emission detected by (\emph{30}). \textbf{(B)} Histograms of the difference between the position angle of the jet and the measured direction of the magnetic field. The green histogram shows the results obtained from the synchrotron emission, whereas the red dashed histogram shows the results obtained by (\emph{30}) from dust polarization.}}

\label{Fig2}  
\end{center} 
\end{figure}

\end{landscape}

\begin{landscape}

\begin{figure}  
\begin{center}

\includegraphics[height=0.8\textwidth]{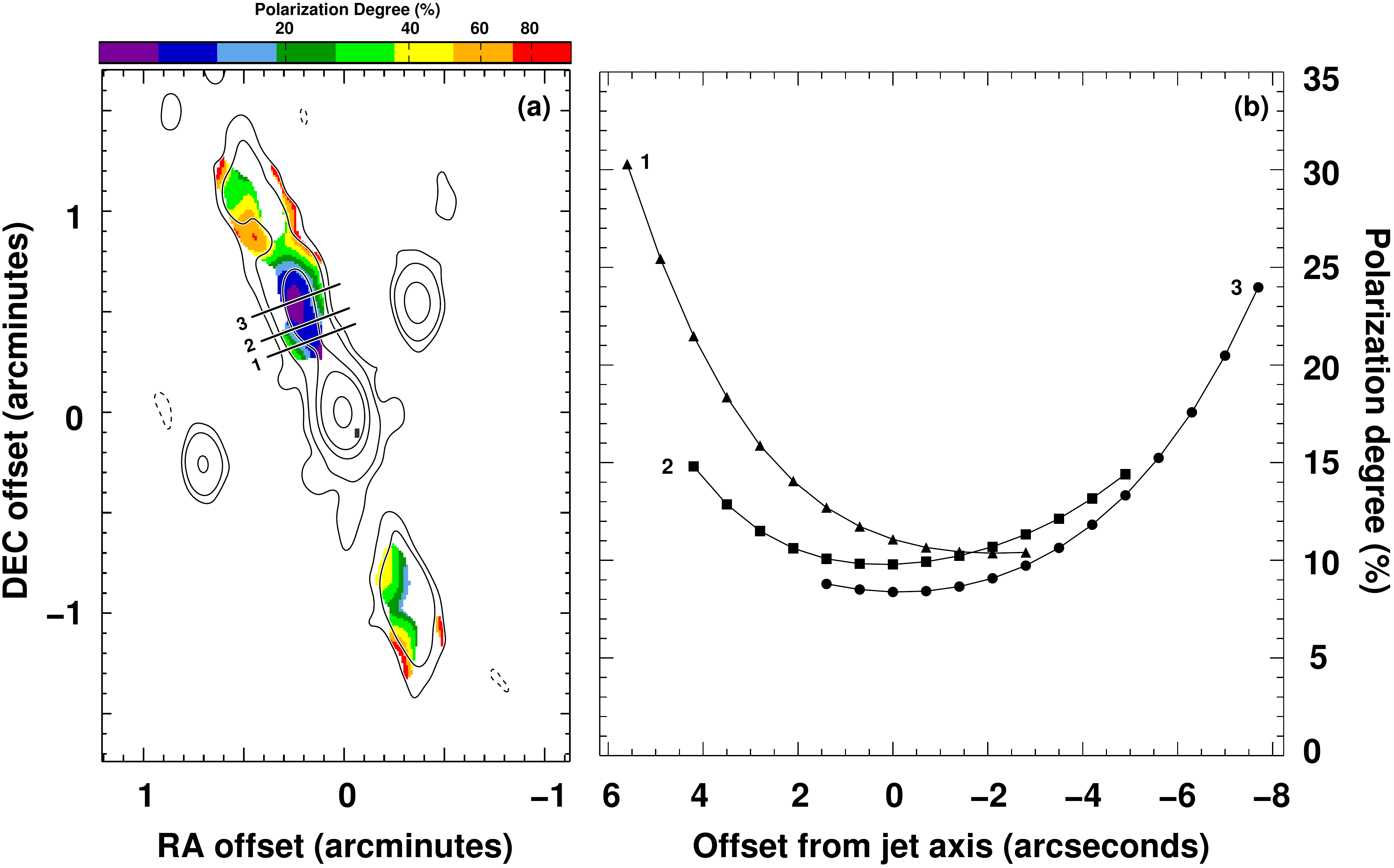}  

\caption{\footnotesize{\textbf{(A)} Image of the continuum emission of the HH~80-81 jet at 6~cm (contours), superposed with the map of polarization degree (colors), obtained as the ratio of the polarized emission over the total continuum emission. Contour levels are 40, 100, 400, 850, and 3300 $\mu$Jy~beam$^{-1}$. The color scale ranges from 5 to 100\%. Black lines mark the direction of the slices shown in (B). \textbf{(B)} Polarization degree for different slices across the jet.}}

\label{Fig3}   
\end{center} 
\end{figure}

\end{landscape}

\newpage

\Large

\noindent \textbf{Supporting Online Material} \\

\large

\noindent \textbf{Materials and Methods} 

\normalsize

\noindent \textbf{Observations and Calibration of the Data}

 Data of continuum emission at 6 cm were obtained using the Very Large Array (VLA) of the National Radio Astronomy Observatory (NRAO). Observations were divided into two runs during 2009 August, 12 and 22. Total on-source time was $\sim$10 hours ($\sim$5 hours in each run). The phase center of the observations was $\alpha$(J2000) = $18^h \, 19^m \, 12.102^s$, $\delta$(J2000) = $-20^\circ \, 47' \, 30.61"$. We used as phase calibrators 1751$-$253, in the first run, and 1924$-$292, in the second run. The amplitude calibrator was 3C~286 in both runs. Data editing and calibration in amplitude, phase and polarization were carried out using the Astronomical Image Processing System (AIPS) package of NRAO, following the standard VLA procedures. 
 
 The polarization calibration was performed using the observations of the amplitude calibrator to determine the absolute polarization angle, while the observations of the phase calibrators were used to determine and correct the antenna-based leakage terms that produce instrumental polarization. The polarization of the phase calibrator will rotate in the sky with parallactic angle while the instrumental polarization will stay constant. We followed the standard VLA procedures described in chapter 4 of the AIPS cookbook.
  
 In order to check the quality of our polarization calibration, we measured the polarization degree and polarization angle of the phase calibrators used in our calibration. We found that these parameters are consistent with those previously found by (\emph{1}) and those tabulated in the VLA Polarization Database for the same calibrators. 
     
 Once calibrated, data from both runs were concatenated using the DBCON task of AIPS. Images of all of the four I, Q, U, and V Stokes parameters were made using the task IMAGR of AIPS. In order to emphasize extended emission, a tapering of 20~k$\lambda$ was used for each image. The rms of the images was 0.013 mJy~beam$^{-1}$, and the synthesized beam was $13"\times8"$ with a position angle of 2$^\circ$. 
 
 We detected emission associated with the HH 80-81 jet in the images of the I, Q, and U Stokes parameters. An image of the total linear polarization emission, $P=\sqrt{Q^2+U^2}$, was obtained using the task COMB of AIPS. Even when the detected polarized emission is relatively weak (3 to 7 $\sigma$), there is an element that gives it additional credibility: the polarized emission is observed only from the non-thermal lobes of the radio jet, with spectral indices of $-$0.8 to $-$0.4, and not from the much brighter central source (the thermal jet) or from the nearby radio sources.

\large

\noindent \textbf{SOM Text} 

\normalsize

\noindent \textbf{Other Possible Emission Mechanisms}

 There are two other mechanisms (besides synchrotron radiation) that could produce the linear polarization observed in the jet of HH~80-81. However, it can be shown that they are too weak to be detected since their optical depths are negligible at radio wavelengths. 

 A first possibility is that we are observing Thomson (i.e. electron) scattering of the radio emission from the central source. This mechanism would produce a polarization pattern similar to that observed. Assuming that part of the radio continuum emission is thermal, optically-thin free-free, and that the region is at a distance of 1.7 kpc (\emph{2}) an upper limit of $n_e\leq10^3$~cm$^{-3}$ is obtained for the electron density in the polarized lobes of the radio jet. From the observed source size of $R\sim3\times10^{17}$~cm ($\sim$12$\arcsec$), and since the Thomson cross section is $\sigma_T=6.65\times10^{-25}$~cm$^2$, we obtain an upper limit of $\tau_T=n_e \, R \, \sigma_T \, \leq \, 2 \times10^{-4}$, for the Thomson opacity, clearly insufficient to produce the observed polarized emission, since an opacity several orders of magnitude higher is required to produce the observed polarized intensity in the lobes of the radio jet. A rough estimate for the radiation dispersed by this process is given by the flux density of the core of the jet ($\sim$ 5 mJy), multiplied by the fraction of solid angle subtended by the scattering region as seen from the core of the jet ($\Omega/4\pi \leq 0.1$) times the Thomson opacity ($2 \times 10^{-4}$). The resulting flux density ($\leq$0.1 $\mu$Jy) is very small and clearly insufficient to be detectable.

 A second mechanism that could produce a polarization pattern similar to that observed in the jet of HH 80-81 is dust dispersion. The region surrounding IRAS~18162$-$2048 has large dust extinction. The H$_2$ column density is $\sim3\times10^{22}$~cm$^{-2}$ (\emph{3}), that translates into an opacity at optical wavelengths of $\tau_V$~$\simeq$~30. However, the cross section for dust scattering scales as $\lambda^{-4}$ (\emph{4}) and its effect is negligible at centimeter wavelengths.

\vspace{1cm}

\begin{table}[hb!]
\begin{center}
\begin{tabular}{cccccccc}
\multicolumn{8}{c}{\textbf{Table S1:} Flux densities and sizes of knots used to} \\
\multicolumn{8}{c}{estimate the magnetic field strength and the energy.} \\
\hline \hline 
     & Source & \multicolumn{3}{c}{Flux Density (mJy)}              & Size         &   B   &       E              \\ \cline{3-5}
Knot &  Name  &     20 cm       &      6 cm       &       2 cm      & (arcseconds) &  (mG) &    (erg)             \\ \hline
N    &   15   & 2.76 $\pm$ 0.14 & 1.62 $\pm$ 0.05 & 1.38 $\pm$ 0.21 &  10          & 0.18  & 1.1$\times$10$^{44}$ \\
S    &   13   & 1.43 $\pm$ 0.18 & 0.95 $\pm$ 0.07 &    ...          &  10          & 0.15  & 7.3$\times$10$^{43}$ \\ \hline
\multicolumn{8}{c}{\textsc{Notes.-} Flux densities and source names from (\emph{5}). The 6 cm flux densities }\\ 
\multicolumn{8}{c}{are consistent with those obtained in our new observations.}
\end{tabular}
\end{center}
\end{table}

\newpage

\begin{center}
\textbf{REFERENCES AND NOTES}
\end{center}

\begin{enumerate}

\item{} Perley, R.A., The positions, structures, and polarizations of 404 compact radio sources. \emph{Astron. J.} \textbf{87}, 859-880 (1983)

\item{} Rodr\'{i}guez, L.F., Moran, J.M., Ho, P.T.P., Gottlieb, E.W., Radio observations of water vapor, hydroxyl, silicon monoxide, ammonia, carbon monoxide, and compact H II regions in the vicinities of suspected Herbig-Haro objects. \emph{Astrophys. J.} \textbf{235}, 845-865 (1980)

\item{} G\'omez, Y., Rodr\'{i}guez, L.F., Girart, J.M., Garay, G., Mart\'{i}, J., VLA and BIMA Observations toward the Exciting Source of the Massive HH 80-81 Outflow. \emph{Astrophys. J.} \textbf{597}, 414-423 (2003)

\item{} Draine, B.~T., Interstellar Dust Grains. \emph{Ann. Rev. Astron.  Astrophys.} \textbf{41}, 241-289 (2003)

\item{} Mart\'{i}, J., Rodr\'{i}guez, L.F., Reipurth, B., HH 80-81: A highly collimated Herbig-Haro complex powered by a massive young star. \emph{Astrophys. J.} \textbf{416}, 208 (1993)

\end{enumerate}

\end{document}